\documentclass[useAMS,usenatbib]{mn2e}
\usepackage{aas_macros,graphicx,times,multirow,amsmath}
\usepackage[]{color}

\title[Swift observations of \src]{\swift\ observations of the ultraluminous X-ray source \src\ in M31}
\author[P.~Esposito et al.] {P.~Esposito,$^{1}$\thanks{E-mail: paoloesp@iasf-milano.inaf.it}  S.~E.~Motta,$^{2}$ F.~Pintore,$^{3,4}$ L.~Zampieri$^{3}$ and L.~Tomasella$^{3}$\smallskip\\
$^1$INAF -- Istituto di Astrofisica Spaziale e Fisica Cosmica - Milano, via E. Bassini 15, I-20133 Milano, Italy\\
$^2$European Space Astronomy Centre (ESAC)/ESA, PO Box 78, E-28691 Villanueva de la Ca\~{n}ada, Madrid, Spain\\
$^3$INAF -- Osservatorio Astronomico di Padova, vicolo dell'Osservatorio 5, I-35122 Padova, Italy\\
$^4$Dipartimento di Astronomia, Universit\`a di Padova, vicolo dell'Osservatorio 3, I-35122 Padova, Italy
}
\date{Accepted 2012 October 14. Received 2012 October 9; in original form 2012 September 14} \pagerange{\pageref{firstpage}--\pageref{lastpage}} \pubyear{2012}

\def\LaTeX{L\kern-.36em\raise.3ex\hbox{a}\kern-.15em
    T\kern-.1667em\lower.7ex\hbox{E}\kern-.125emX}

\def\xmm {\emph{XMM-Newton}}
\def\cxo {\emph{Chandra}}
\def\swift {\emph{Swift}}

\def\src {XMMU\,J004243.6+412519}
\def\flux {\mbox{erg cm$^{-2}$ s$^{-1}$}}
\def\lum {\mbox{erg s$^{-1}$}}
\def\nh {$N_{\rm H}$}

\begin{document}

\label{firstpage}
\maketitle
\begin{abstract}
We report on a multi-wavelength study of the recently discovered X-ray transient \src\ in M31, based on data collected with \swift\ and the 1.8-m Copernico Telescope at Cima Ekar in Asiago (Italy) between 2012 February and August. Undetected in all previous observations, in 2012 January \src\ suddenly turned on, showing powerful X-ray emission with a luminosity of $\sim$$10^{38}$ \lum\ (assuming a distance of 780 kpc). In the following weeks, it reached a luminosity higher than $\sim$$10^{39}$ \lum, in the typical range of ultraluminous X-ray sources (ULXs).  For at least $\sim$40 days the source luminosity remained fairly constant, then it faded below $\approx$$10^{38}$ \lum\ in the following $\sim$200 days. The source spectrum, which can be well described by multi-color disk blackbody model, progressively  softened during the decay (the temperature changed from $kT\sim0.9$ keV to $\sim$0.4 keV). No emission from \src\ was  detected down to 22 mag in the optical band and to 23--24 mag in the near ultraviolet. We compare the properties of \src\ with those of other known ULXs and Galactic black hole transients, finding more similarities with the latter.

\end{abstract}

\section{Introduction}
\label{sect1}

Ultra luminous X-ray sources (ULXs; \citealt{long83,h84,fabbiano89}) are observed in nearby galaxies as point-like, off-nuclear X-ray sources with isotropic luminosity larger than the Eddington limit for spherical accretion of fully ionised hydrogen onto a $\sim$10-$M_\odot$ compact object ($\approx$$10^{39}$ \lum). The nature of these objects remains a major astrophysical puzzle. Some ULXs have been identified with X-ray bright supernovae, other with background active galactic nuclei, but most of them are suspected to be accreting black holes (BHs) of stellar origin (see \citealt{feng11} for a recent review). However, their luminosities are typically too high to be explained by accretion processes on stellar-mass black holes without violating standard Eddington limit arguments. Three main different possibilities have been proposed to explain the nature of ULXs. (1) They are intermediate-mass BHs (with mass in excess of 100 $M_\odot$) in accretion \citep{colbert99}; a strong case for the existence of such objects is HLX-1 in ESO 243-49, for which the
luminosity exceeds $10^{42}$ \lum\ \citep{farrell09,davis11}. (2) They are accreting stellar-mass BHs ($\sim$5--20 $M_{\odot}$) with substantial beaming of their X-ray emission \citep{king01} or emitting above their Eddington limit \citep*{begelman02,begelman06}. (3) They are massive BHs ($\sim$30--80 $M_{\odot}$) formed in low-metallicity environments from massive progenitors  and accreting at or slightly above their Eddington limit (\citealt*{mapelli09}; \citealt{zampieri09,belczynski10}). Other possibilities are that the anomalously high luminosity of these objects is due to a combination of these factors, or that the population of ULXs is composed by different kind of sources.

Observations of ULXs' counterparts in energy bands other than X-ray have proven difficult because of their extragalactic nature. In X-rays, ULXs have been extensively observed during the last decade with high-spectral-resolution satellites such as \textit{Chandra} and \textit{XMM-Newton}. High-quality spectra taken with such satellites have shown that many bright persistent sources emit a spectrum commonly well fit by a Comptonization model plus a disc (\citealt*{stobbart06,roberts07,gladstone09}). These spectral properties appear consistent with super-Eddington or marginally super-Eddington accretion onto a stellar-mass or a more massive BH (e.g. \citealt*{gladstone09,pintore12}). Quasi-periodic oscillations (QPOs) similar to the ones observed in the emission of Galactic BH binaries (see e.g. \citealt*{casella05}) have been detected in X-rays in a few ULXs (\citealt{mucciarelli06,strohmayer07}; \citealt*{sm03,strohmayer09,feng10}). Some efforts have been made to make use of scaling arguments valid for Galactic BH binaries (see e.g. \citealt{shaposhnikov09}) in order to constrain the masses of ULXs from characteristic time-scales.  Results are, however, highly uncertain, as all these estimates are based upon tentative identifications of the timing features and the use of scaling laws that are known to hold only for a limited number of objects (e.g. \citealt*{middleton11}).

Despite the substantial body of work that has been done in the last decade, transient ULXs in particular still represent a poorly know population. Only few ULXs with transient behaviour have been detected so far and only a handful of them have been studied in detail. Examples of well-studied genuine ULX transients are CXOM31\,J004253.1+411422 in M31 (hereafter M31 ULX-1), CXOU\,133705.1--295207  in M83 and CXOU\,J132518.2--430304 in NGC5128 \citep{kaur12,soria12,sivakoff08short}. In 2012 January a new X-ray source with a luminosity of $\sim$$10^{38}$ \lum\ was discovered by \xmm\ in M31 (\src; \citealt*{henze12}). Seven days after its discovery, it reached a luminosity of $\sim$ $2\times10^{39}$ \lum, which made it the second most luminous ULX in M31 \citep*{hph12}. During the outburst the source was repeatedly observed by \swift\ and other instruments. Here we present the analysis of the several \swift\ XRT and UVOT observations performed between 2012 March and August with the aim to explore the nature of this source. We also present the results of deep optical observations taken on 2012 July 18 with the 1.8-m Copernico Telescope at Cima Ekar  (Asiago, Italy) and the analysis of serendipitous observations of the source region collected in recent years by \swift\ and \cxo.

In Section~\ref{obsdata} we describe the \swift\ observations used in our study and we present the results of the spectral and timing analysis of the X-ray data in Section~\ref{xrayspectra}. In Section~\ref{uvoptical} we report on the \swift/UVOT and Copernico optical and ultraviolet observations of the source. In Section~\ref{archivaldata} we present the upper limits on the pre-outburst X-ray flux of \src\ obtained from our inspection of \cxo\ and \swift\ archival observations. Discussion follows in Section~\ref{disc}.

\section{\swift\ observations and data reduction}\label{obsdata}

The \swift\ payload includes a wide-field instrument, the coded-mask gamma-ray Burst Alert Telescope \citep[BAT;][]{barthelmy05short}, and two narrow-field instruments, the X-Ray Telescope \citep[XRT;][]{burrows05short} and the Ultra-Violet/Optical Telescope \citep[UVOT;][]{roming05short}. In this work we made use of only the narrow-field instruments data.

The XRT uses a front-illuminated CCD detector sensitive to photons between 0.2 and 10 keV. Two main readout modes are available: photon counting (PC) and windowed timing (WT). PC mode provides two dimensional imaging information and a 2.507-s time resolution; in WT mode only one-dimensional imaging is preserved, achieving a time resolution of 1.766 ms. The UVOT is a 30-cm modified Ritchey-Chr\'etien reflector using a microchannel-intensified CCD detector which operates in photon counting mode. A filter wheel accommodates a set of optical and ultraviolet (UV) filters and the wavelength range is 1700--6000 \AA. The data were processed and filtered with standard procedures and quality cuts\footnote{See http://swift.gsfc.nasa.gov/docs/swift/analysis/ for more details.} using \textsc{ftools} tasks in the \textsc{heasoft} software package (v.~6.12) and the calibration files in the 2012-02-06 \textsc{caldb} release. 

Following the discovery of \src\ (2012 January; \citealt*{henze12}), the source was observed by \swift\ ten times in about three weeks, until it came out of visibility in 2012 March. At the end of 2012 May, \src\  became visible again for \swift\ and the monitoring was resumed with many further pointings, mostly off-axis (see also \citealt{hphg12}). Around mid 2012 August, the source flux became too low for the typical sensitivity of a $\sim$2-ks XRT snapshot. For this reason, after the visibility gap we consider only the sixteen observations taken up to 2012 September 01. A summary of the observations used in this work is given in Table~\ref{obs-log}.

\begin{table*}
\begin{minipage}{11.cm}
\centering \caption{\swift\ observations used for this work.} \label{obs-log}
\begin{tabular}{@{}lccccc}
\hline
Observation & XRT  & UVOT  & \multicolumn{2}{c}{Start / end time (UT)} & Exposure\\
 & mode & filter$^{a}$ & \multicolumn{2}{c}{(YYYY-MM-DD hh-mm-ss)} & (ks)\\
\hline
00032286002 & PC & $u$ & 2012-02-19 00:47:34 & 2012-02-19 23:23:57 & 3.9 \\
00032286003 & PC & $u$ & 2012-02-23 18:49:39 & 2012-02-23 22:13:56 & 3.3 \\
00032286004 & WT & -- & 2012-02-24 13:54:21 & 2012-02-24 15:35:02 & 0.5 \\
00032286005 & WT & $uvw2$ & 2012-02-28 04:50:00 & 2012-02-28 08:17:00 & 3.0 \\
00032286009 & WT & $uvw2$ & 2012-03-02 03:02:59 & 2012-03-02 03:45:40 & 2.5 \\
00032286006 & PC & $u$ & 2012-03-02 04:54:08 & 2012-03-02 14:58:57 & 6.3 \\
00032286007 & PC & $uvw2$ & 2012-03-03 00:09:07 & 2012-03-03 16:37:58 & 7.1 \\
00032286008 & WT & $uvw2$ & 2012-03-03 01:41:44 & 2012-03-03 06:57:00 & 2.7 \\
00032286010 & PC & $uvm2$ & 2012-03-04 04:52:27 & 2012-03-04 21:21:57 & 4.3 \\
00032286011 & WT & $uvw2$ & 2012-03-07 10:00:31 & 2012-03-07 12:01:00 & 3.1 \\
\hline
00035336052 & PC & $uvw1$ & 2012-05-24 14:55:42 & 2012-05-24 21:36:56 & 4.3 \\
00032286012 & WT & $uvw1$ & 2012-05-28 11:55:26 & 2012-05-28 15:37:59 & 4.1 \\
00035336053 & PC & -- & 2012-06-01 18:28:39 & 2012-06-01 23:43:57 & 4.0 \\
00035336054 & PC & $uvw1$ & 2012-06-09 06:20:09 & 2012-06-09 11:22:57 & 2.0 \\
00035336055 & PC & -- & 2012-06-17 03:17:31 & 2012-06-17 08:14:56 & 2.1 \\
00035336056 & PC & $uvw1$ & 2012-06-25 00:42:19 & 2012-06-25 18:40:57 & 1.9 \\
00035336058 & PC & $uvw1$ & 2012-07-09 16:01:04 & 2012-07-09 19:30:55 & 1.1 \\
00035336059 & PC & -- & 2012-07-11 06:36:01 & 2012-07-11 06:37:49 & 0.1 \\
00035336060 & PC & $uvw1$ & 2012-07-15 11:37:32 & 2012-07-15 11:55:55 & 1.1 \\
00035336061 & PC & $uvw1$ & 2012-07-19 02:11:18 & 2012-07-19 15:23:56 & 2.1 \\
00035336062 & PC & $uvw1$ & 2012-07-27 18:33:47 & 2012-07-27 20:25:54 & 2.3 \\
00035336063 & PC & -- & 2012-08-05 03:03:49 & 2012-08-05 11:19:53 & 2.0 \\
00035336064 & PC & -- & 2012-08-12 11:32:21 & 2012-08-12 21:20:56 & 0.9 \\
00035336065 & PC & $uvw1$ & 2012-08-20 21:29:05 & 2012-08-20 23:24:55 & 2.1 \\
00035336066 & PC & $uvw1$ & 2012-08-28 13:49:33 & 2012-08-28 14:05:53 & 1.0 \\
00035336067 & PC & $uvw1$ & 2012-09-01 13:58:43 & 2012-09-01 14:24:55 & 1.6 \\

\hline
\end{tabular}
\begin{list}{}{}
\item[$^{a}$] $u$:  central wavelength 3465 \AA, FWHM 785 \AA; $uvw1$:  central wavelength 2600 \AA, FWHM \mbox{693 \AA}; $uvm2$:  central wavelength 2246 \AA, FWHM \mbox{498 \AA}; $uvw2$:  central wavelength 1928 \AA, FWHM 657 \AA.
\end{list}
\end{minipage}
\end{table*}

\section{X-ray data analysis}\label{xrayspectra}

We extracted the PC source events from a circle with a radius of 20 pixels (1 XRT pixel corresponds to about $2\farcs36$) and the WT data from a $40\times40$ pixels box along the image strip. To estimate the background, we extracted PC and WT events from regions far from the position of \src. The ancillary response files (arf) were generated with \textsc{xrtmkarf}, and they account for different extraction regions, vignetting and point spread function corrections. We used the latest available spectral redistribution matrix (rmf) in \textsc{caldb}. The spectral channels were grouped so as to have bins with a minimum number of 20 photons. The spectral analysis was performed with the \textsc{xspec} 12.7 fitting package; \citep{arnaud96}; the abundances used are those of \citet{anders89} and photoelectric absorption cross-sections are from \citet{balucinska92}.

\begin{figure}
\resizebox{\hsize}{!}{\includegraphics[angle=-90]{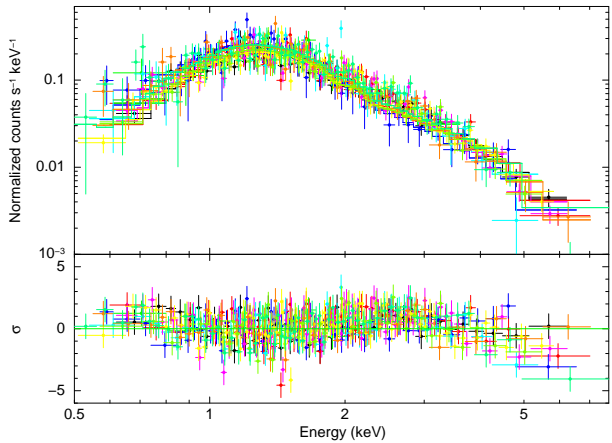}}
\caption{\label{allspec} Simultaneous modelling of all the 2012 February--March PC and WT spectra using an absorbed power-law model. Bottom panel: the residuals of the fit (in units of standard deviations). (See the electronic journal for a colour version of this figure.)}
\end{figure}

Initially we focus our analysis on the 2012 February--March data (see Table~\ref{obs-log}). For a preliminary look at the data, we fit all spectra simultaneously (in the 0.5--10 keV energy range) with the hydrogen column density tied between all observations using a simple power law (e.g. \citealt*{hph12}). While this simultaneous modelling yields a rather high reduced $\chi^2$ of 1.37 for 553 degrees of freedom (dof), the test shows that, as can be seen in Fig.~\ref{allspec}, the spectra of \src\ are all very similar in this set of XRT observations (see also \citealt*{hepiha12}).

Thus, in order to achieve better statistics and higher signal-to-noise ratio, we merged the data from all the observations and accumulated combined PC and WT spectra. In the following, we concentrate on the PC total spectrum (24.9 ks, about 7900 counts, 99.4\% of which are attributable to \src), because of the intrinsically higher signal-to-noise ratio of the PC data with respect to the WT ones. The combined spectrum was fit adopting several phenomenological models frequently used in literature for ULXs (e.g. \citealt{roberts07,feng11}): a \textsc{diskbb}, a \textsc{powerlaw},  a \textsc{diskbb+powerlaw}, and a \textsc{diskbb+comptt}, all corrected for interstellar absorption. All these models but the power law provide statistically good fits. However, the improvement given by the combination of two models is not significant in comparison with a single \textsc{diskbb} model ($\chi_\nu^2=1.07$ for 231 dof against, for instance, $\chi_\nu^2=1.03$ for 229 dof for the \textsc{diskbb+powerlaw}). We also tried to fit more sophisticated Comptonization models (\textsc{comptt} and \textsc{simpl} in \textsc{xspec}) to the data, but  the count statistics of the spectrum is not sufficient to provide good-enough constraints on the model parameters. Hence, it seems appropriate to describe the pre-gap PC spectrum of \src\ in terms of a simple \textsc{diskbb} model. The best fitting parameters show that the disc is relatively hot ($0.86 \pm 0.02$ keV) and the luminosity is quite high ($\sim$$1.1\times10^{39}$ \lum\  for a distance of 780 kpc; \citealt{holland98,stanek98}). The results are summarised in Table~\ref{fits}. 
\begin{table*}
\centering 
\scalebox{0.88}{\begin{minipage}{19.9cm}
\caption{Spectral analysis of \src\ (2012 February--March, PC data). Errors are at a 1$\sigma$ confidence level for a single parameter of interest.} \label{fits}
\begin{tabular}{@{}lcccccccc}
\hline
Model$^{a}$ & \nh & $\Gamma_1$ & $E_{\mathrm{cut/break}}$ & $kT_{in}$ & $\Gamma_2$ / Norm.$^b$ & Observed flux$^c$ & Luminosity$^c$ & $\chi^2_\nu$ (dof)\\
 & ($10^{21}$ cm$^{-2}$)& & (keV) & (keV) & & ($10^{-11}$ \flux) &($10^{39}$ \lum) \\
\hline
\textsc{phabs*powerlaw} & $6.7\pm0.2$& $2.91\pm0.04$ & -- & -- & -- & $1.12\pm0.02$ & $2.33\pm0.09$ & 1.46 (231)\\
\textsc{phabs*diskbb} & $2.9\pm0.1$& -- & -- & $0.86\pm0.02$ & $1.4\pm0.1$ & $1.04\pm0.02$ & $1.04^{+0.02}_{-0.01}$ & 1.07 (231)\\
\textsc{phabs*cutoffpl} & $3.5\pm0.3$& $0.7^{+0.2}_{-0.3}$ & $1.4^{+0.1}_{-0.2}$ & -- & -- & $1.06^{+0.02}_{-0.01}$ &$1.15^{+0.07}_{-0.06}$ & 1.03 (230)\\
\textsc{phabs*bknpower} & $4.5\pm0.3$& $2.1^{+0.1}_{-0.2}$ & $2.6^{+0.2}_{-0.1}$ & -- & $3.6^{+0.2}_{-0.1}$ & $1.10\pm0.02$ &$1.44^{+0.09}_{-0.08}$ & 1.00 (229)\\
\textsc{phabs*(powerlaw+diskbb)} & $3.2^{+0.4}_{-0.2}$& $1.6^{+0.6}_{-1.8}$ & -- & $0.79^{+0.04}_{-0.02}$ & $1.8\pm0.2$ & $1.10\pm0.02$ & $1.12^{+0.03}_{-0.02}$ & 1.03 (229)\\
\textsc{phabs*(cutoffpl+diskbb)} & $9^{+1}_{-2}$& $1.4\pm0.5$ & $1.7^{+0.5}_{-0.3}$ & $0.12^{+0.02}_{-0.01}$ & $1.1^{+4.3}_{-0.9}\times10^5$ & $1.07^{+0.01}_{-0.02}$ & $1.15^{+0.07}_{-0.06}$ & 1.00 (228)\\
\hline
\end{tabular}
\begin{list}{}{}
\item[$^{a}$] \textsc{xspec} model.
\item[$^{b}$] \textsc{diskbb} normalisation: $(R[\mathrm{km}]/D[10~\mathrm{kpc}])^2\cos \theta$, $\theta=0$ corresponding to a face-on disk.
\item[$^{c}$] In the 0.5--10 keV energy range; for the luminosity we assumed a distance to M31 of 780 kpc \citep{holland98,stanek98}.
\end{list}
\end{minipage}}
\end{table*}

\begin{figure}
\resizebox{\hsize}{!}{\includegraphics[angle=-90]{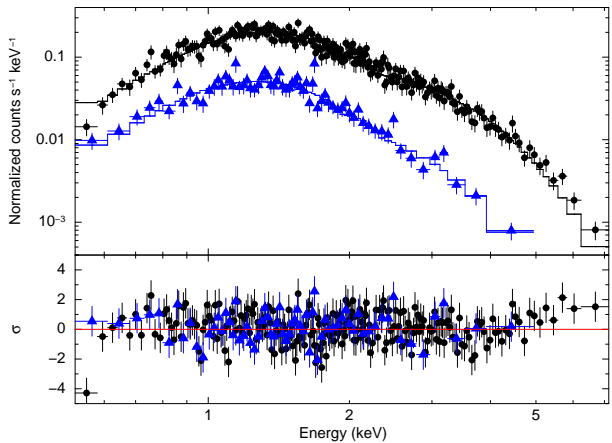}}
\caption{\label{spec_compare} Comparison of the pre- (black circles) and post-visibility gap (blue triangles) cumulative PC spectra for the \textsc{phabs*diskbb} model (see Section~\ref{xrayspectra} for details). Bottom panel: the residuals of the fit (in units of standard deviations).}
\end{figure}

\begin{figure}
\resizebox{\hsize}{!}{\includegraphics[angle=-90]{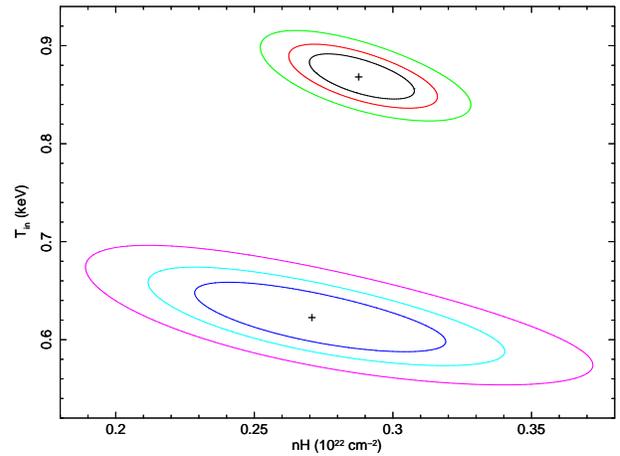}}
\caption{\label{cplot} Contour plots of temperature versus column density (adopting the \textsc{phabs*diskbb} model) for the pre-gap (black/red/green upper contours) and post-gap (blue/cyan/violet lower contours) PC data. The crosses indicate the best fits. Lines mark 1, 2 and 3$\sigma$ confidence levels.}
\end{figure}

Similarly, for the second batch of data (collected starting from 2012 May; see Table~\ref{obs-log}) we extracted a cumulative PC spectrum (23.7 ks, about 1600 counts, 99.2\% of which are attributable to \src). We tested the same single-component\footnote{Owing to the lower count statistics, this time we did not consider more complicated (two-component) models.} spectral models reported in Table~\ref{fits}, obtaining the parameters shown in Table~\ref{fits2}. In general, they appear to have values of the column density consistent with those of the pre-gap state and significantly lower fluxes/luminosities. In particular the post-gap average spectrum is well described by a \textsc{diskbb} component with an (average) temperature ($\sim$0.6 keV) smaller than that of the pre-gap combined spectrum. In Fig.~\ref{spec_compare} we show the \textsc{phabs*diskbb} model fit  to the pre- and post-visibility gap combined PC spectra. The softening suggested by the spectral parameters and Fig.~\ref{spec_compare} is apparent from the contour plots shown in Fig.~\ref{cplot}.
\begin{table*}
\centering 
\scalebox{0.88}{\begin{minipage}{19.9cm}
\caption{Spectral analysis of \src\ (2012 May--August, PC data). Errors are at a 1$\sigma$ confidence level for a single parameter of interest.} \label{fits2}
\begin{tabular}{@{}lcccccccc}
\hline
Model$^{a}$ & \nh & $\Gamma_1$ & $E_{\mathrm{cut}}$ & $kT_{in}$ & $\Gamma_2$ / Norm.$^b$ & Observed flux$^c$ & Luminosity$^c$ & $\chi^2_\nu$ (dof)\\
 & ($10^{21}$ cm$^{-2}$)& & (keV) & (keV) & & ($10^{-12}$ \flux) &($10^{39}$ \lum) \\
\hline
\textsc{phabs*powerlaw} & $6.9\pm0.5$& $3.6\pm0.1$ & -- & -- & -- & $2.04\pm0.07$ & $0.71^{+0.10}_{-0.08}$ & 1.45 (66)\\
\textsc{phabs*diskbb} & $2.7\pm0.3$& -- & -- & $0.62\pm0.02$ & $1.2\pm0.2$ & $1.95\pm0.06$ & $0.22\pm0.01$ & 0.98 (66)\\
\textsc{phabs*cutoffpl} & $1.5^{+1.0}_{-0.8}$ & $-1.0^{+0.8}_{-0.6}$ & $0.6\pm0.1$ & -- & -- & $1.95\pm0.06$ & $0.17^{+0.04}_{-0.02}$ & 0.98 (65) \\
\textsc{phabs*bknpower} & $3.8^{+0.7}_{-0.9}$ & $2.2^{+0.3}_{-0.5}$ & $2.3^{+0.1}_{-0.3}$ & -- & $4.6^{+0.4}_{-0.5}$ & $2.01^{+0.06}_{-0.07}$ & $0.29\pm0.05$ & 0.99 (64) \\
\hline
\end{tabular}
\begin{list}{}{}
\item[$^{a}$] \textsc{xspec} model.
\item[$^{b}$] \textsc{diskbb} normalisation: $(R[\mathrm{km}]/D[10~\mathrm{kpc}])^2\cos \theta$, $\theta=0$ corresponding to a face-on disk.
\item[$^{c}$] In the 0.5--10 keV energy range; for the luminosity we assumed a distance to M31 of 780 kpc \citep{holland98,stanek98}.
\end{list}
\end{minipage}}
\end{table*}

\begin{figure}
\resizebox{\hsize}{!}{\includegraphics[angle=-90]{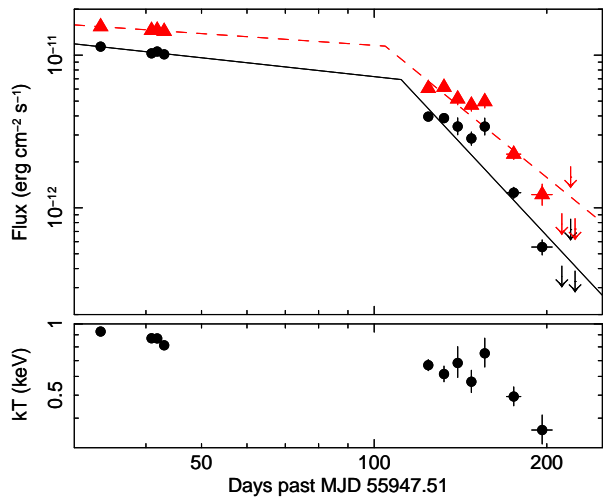}}
\caption{\label{decay}The top panel shows time evolutions of the absorbed (black circles) and unabsorbed (red triangles) fluxes in the 0.5--10 keV energy range for the \textsc{phabs*diskbb} model (see Section~\ref{xrayspectra} for details); the down-arrows indicate upper limits at the 3$\sigma$ confidence level. The broken-power-law models describing the decays are also plotted (black solid line for the observed flux and red dashed line for the unabsorbed flux). Bottom panel: evolution of the characteristic temperature of the \textsc{diskbb} model inferred from the spectral fitting. We assumed as $t = 0$ the date the source was observed for the first time to exceed the ULX threshold \citep*{hph12}.}
\end{figure}

In order to study the source flux evolution over time, we fit all the PC spectra from the individual observations or from small groups of observations. When the source flux dropped below $2\times10^{-12}$ \flux, around mid 2012 July, in order to accumulate sufficient statistics for meaningful spectral fits, we combined data from a few contiguous observations; namely, we obtained a spectrum from observations 00035336058--61 and another one from observations 00035336062--64.  We fit all the spectra simultaneously adopting the \textsc{phabs*diskbb} model with the hydrogen column density (which is consistent with a single value in both the pre- and post-gap cumulative spectra) tied between observations ($\chi^2_\nu=1.10$ for 676 dof). We plot the resulting long-term light curve and the inferred temperatures $kT$ in Fig.~\ref{decay}. We show both the absorbed and unabsorbed fluxes. As can be seen from Fig.~\ref{decay} the effect of the interstellar absorption is larger on the softer spectra from the post-gap observations. Progressive spectral softening and flux decay are evident during the post-gap observations. The flux decreased by a factor of $\approx$5 over $\sim$70 days (by a factor of $\approx$10  over $\sim$150 days with respect to the pre-gap flux), while the disk temperature changed from $kT\sim0.7$ keV to $\sim$0.4 keV.

Although, owing to the long visibility gap, the available data do not allow us to perform an accurate modelling of the decay shape, we tried a number of simple models to fit the light curve of \src. We fixed $t=0$ at the time the source was observed for the first time to exceed the ULX threshold (MJD 55947.51; \citealt*{hph12}). An exponential function of the form $F(t) = A \exp(-t/\tau)$ gives a rather poor fit ($\chi^2_\nu=11.58$ for 10 dof). The best-fitting parameters for the observed (absorbed) flux are $A=(2.00\pm0.05)\times10^{-11}$ \flux\ and $e$-folding time $\tau=(63.9\pm1.3)$ d.

Among the other models tested, a broken-power-law model provides the better fits for the evolution of both the absorbed  ($\chi^2_\nu=4.33$ for 8 dof) and unabsorbed fluxes ($\chi^2_\nu=3.65$ for 8 dof).  Assuming as $t=0$ MJD 55947.51, for the observed flux, the break occurs at ($111.4\pm6.4$) d, when the index changes from $\alpha_1 = - 0.41\pm0.18$ to $\alpha_1 = - 4.03\pm0.19$; the flux at the break time is $(6.9\pm1.5)\times10^{-12}$ \flux. For the unabsorbed flux, the best-fitting parameters are: $\alpha_{\circ,1} = - 0.26\pm0.15$, $\alpha_{\circ,2} = - 3.06\pm0.19$ and $(104.5\pm6.4)$ d for the break epoch (with an unabsorbed flux of $(1.1\pm0.2)\times10^{-11}$ \flux). The steeper decay of the absorbed flux is due to the fact that the absorption affects the softer (later) spectra more. We stress that the fit parameters depend on the assumed time origin. The uncertainties introduced by making different (reasonable) assumptions can be larger than the statistical errors reported here. For instance, assuming $t=0$ at MJD 55942 (the day after the discovery of the source at a luminosity of $\approx$ $2\times10^{38}$ \lum; \citealt*{henze12}), we find for the observed flux: $\alpha_{1}'= - 0.19\pm0.12$, $\alpha_{2}' = - 4.18\pm0.19$, $(110.3\pm3.6)$ d for the break epoch and flux at the break time of $(8.8\pm1.0)\times10^{-12}$ \flux. 

In the last observations (segments 00035336065, 00035336066 and 00035336067), \src\ was not detected. The $3\sigma$ upper limits on the XRT count rate derived from the deeper observations (6065 and 6067) are of $\sim$0.1 counts s$^{-1}$ (0.3--10 keV, following \citealt*{kraft91}). Assuming the spectrum of the closest observations (6062--6064, $kT\simeq0.4$ keV), this corresponds to upper limits on the observed flux of $\sim$$4\times10^{-13}$ \flux\ and  of $\sim$$6\times10^{37}$ \lum\ on the luminosity (for 780 kpc).

For the timing analysis we concentrate on the data in WT mode (total exposure: 16.0 ks; see Table~\ref{obs-log}), since their time resolution of $\sim$1.7 ms (corresponding to a Nyquist frequency of $\sim$280 Hz) makes it possible to search for fast time variability. For each observation we computed a power density spectrum (PDS) in the energy band 0.3--10 keV by using intervals up to $\sim$925-s long and averaging the individual spectra for each observation. The PDS did not reveal any significant variability from \src: neither broad-band components nor narrow features were detected; no significant evolution of the PDS in time could be detected as well. 

To improve the statistics, we produced a single PDS averaging all the WT data. The PDS was normalised according to \citet{leahy83}, so that powers due to Poissonian counting noise have an average value of 2. Since PDS created from data taken in WT mode show a drop-off at high frequencies\footnote{This effect is mostly visible above $\sim$ 50 Hz  and is related to the read-out method of the XRT detector, see http://www.swift.ac.uk/analysis/xrt.} and since also the averaged PDS from all the WT observations did not show any significant feature, we rebinned our data at a Nyquist frequency of $\sim$17 Hz (rebin factor of 16) using data stretches 925-s long to obtain a new average PDS. The PDS is well fit by a constant component equal to 2.286 $\pm$ 0.006 (reasonably describing the Poissonian noise) with a best-fitting $\chi ^2$ = 169.56 for 152 dof. Hence we conclude that no significant variability is detected from the source. It is not possible, however, to exclude that the emission from the source has a certain level of variability hidden by photon counting statistics or on time-scales not accessible to our data ($\la$$10^{-4}$ Hz). 

Apart from the flux decay on the scale of weeks, the emission from \src\ does not show signs of strong aperiodic variability either. There is only some evidence of moderate variability on the 1-ks-scale, with a rms variance of $(20\pm5)$ percent in the WT data. This is probably related to some level of variability of the local astrophysical background and could account for the fact that the Poissonian noise level observed in the PDS is slightly higher than the expected value (2 with the Leahy normalisation). For the pre-gap PC light curves (a similar analysis of the post-gap data is hampered by the low count rates) the 3$\sigma$ upper limits on the rms variability range from approximately 12 percent to 20 percent.

\section{Optical and ultraviolet observations}\label{uvoptical}

The \swift/UVOT observed \src\ simultaneously with the XRT. The data were taken with the $u$, $uvw1$, $uvm2$ and $uvw2$ filters (see Table~\ref{obs-log}). The analysis was performed on the individual and stacked (for each filter) images with the \textsc{uvotsource} task, which calculates the magnitude through aperture photometry within a circular region (we used a 3-arcsec radius) and applies specific corrections due to the detector characteristics.

No source was detected at the position of \src\ in any of the UVOT observations and filters, before or after the visibility gap. The 3$\sigma$ limits before the visibility gap are in the stacked images $u>23.2$ mag (total exposure: 13.4 ks), $uvm2>23.7$ mag (total exposure: 5.6 ks) and $uvw2>24.5$ (total exposure: 18.2 ks). After the gap, \src\ was observed only with the $uvw1$ filter; the 3$\sigma$ limit from the stacked image is $uvw1>24.4$ (total exposure: 23.2 ks) All magnitudes are in the AB system \citep{oke83}; see \citet{poole08short} for more details on the UVOT photometric system and \citet{breeveld11} for the most updated zero-points and count rate to flux conversion factors. 

\begin{figure}
\resizebox{\hsize}{!}{\includegraphics[angle=0]{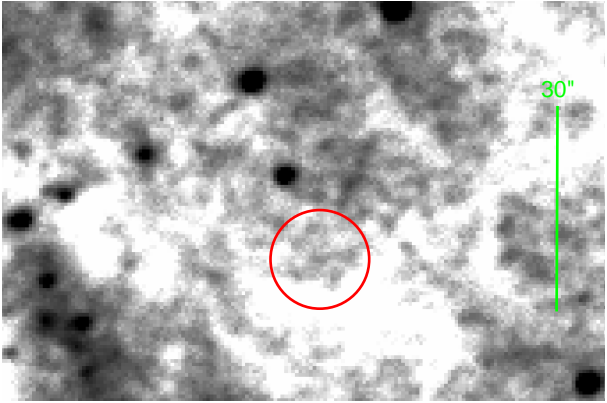}}
\caption{\label{asiago} $V$-band image of the field of \src\ taken on 2012 July 18 at the 1.8-m Copernico Telescope at Cima Ekar in Asiago. North is up, East to the left. The green circle ($7\farcs2$, 3$\sigma$) is centred on the \cxo\ position of \src\ \citep{barnard12}.}
\end{figure}

The above magnitudes have not been corrected for extinction. At the position of \src, the total line-of-sight optical extinction estimated from background infrared emission is  $A_V=0.05$ mag \citep*{schlegel98}, while the X-ray fits, adopting the relation $N_{\mathrm{H}}=1.79\times10^{21}A_V$ cm$^{-2}$  by \citet{predehl95}, yield higher values in the range $A_V\approx1.7$--5.0. Indicatively, $A_V=1$ mag corresponds to $A_{u}\simeq1.8$ mag, $A_{uvw1}\simeq2.2$ mag , $A_{uvm2}\simeq3.2$ mag, and $A_{uvw2}\simeq2.8$ mag  \citep{fitzpatrick07}.

The field of \src \ was also observed with the 1.8-m Copernico Telescope at Cima Ekar in Asiago (Italy) on 2012 July 18. Three images of 20 minutes were taken in both the $V$ and $B$ band filters. The data were reduced following standard prescriptions. After removal of the detector signature (bias and flat field corrections), the images were astrometrically calibrated performing a polynomial interpolation starting from the positions of the NOMAD star catalogue \citep{zacharias05}. The accuracy is $0\farcs2$. The three calibrated frames in each filter were then averaged and the resulting $V$ band image is shown in Fig.~\ref{asiago}. 

Instrumental magnitudes were measured on the images through the point-spread function (PSF) fitting technique. The photometric calibration was performed using reference stars from the catalogue of M31 compiled by \cite{magnier92}, homogeneously distributed around the source position. The internal accuracy of this calibration is 0.1 mag in both bands. No source at the position of \src\ was detected in any of the two filters down to a limiting magnitude of 21.7 and 22.2 in the $V$ and $B$ band filters, respectively (see Fig.~\ref{asiago}). In fact, the background emission from M31 is highly variable inside the error box, so that the actual limiting magnitude varies from 21.5 to 21.9 in $V$ and from 21.6 to 22.8 in $B$, depending on the position.

\section{Pre outburst observations}\label{archivaldata}
As noted by \citet*{henze12,hph12}, no source compatible with the position of \src\ was listed in any X-ray catalogue. We have searched the \xmm, \cxo, and \swift\ public archives for possible previous bright states of \src\ in recent years, but the source was never detected. Since the \xmm\ observations are rather sparse, in the following we summarise only the upper limits obtained from the much more intense coverage (from 1999 November to 2012 January) with \cxo\ and \swift.

\subsection{\cxo}
\begin{figure}
\resizebox{\hsize}{!}{\includegraphics[angle=-90]{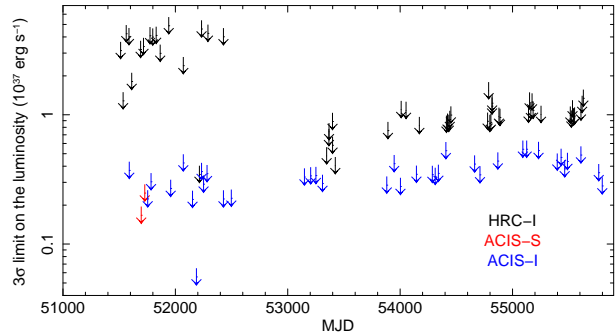}}
\caption{\label{ulcxo} \cxo\ 3$\sigma$ upper limits on the 0.5--10 keV luminosity (for a distance of 780 kpc; \citealt{holland98,stanek98}). (See the electronic journal for a colour version of this figure.)}
\end{figure}

The distribution of the 93 \cxo\ observations (from 1999 November 30 to 2011 August 25) covering the field of \src\ can be seen in Fig.~\ref{ulcxo}; 54 observations were carried out with the HRC-I instrument \citep{murray00}, 39 with the ACIS (I or S; \citealt{garmire03}).  Typical exposures are for the HRC-I  $\sim$1--5 ks in the 1999--2011 observations and $\sim$20 ks in the more recent ones, and $\sim$5 ks for the ACIS pointings; the deepest upper limit (see Fig.~\ref{ulcxo}) was obtained from a 38-ks ACIS-I observation carried out on 2001 October 05 (obs. ID: 1575; MJD 52187). 

For each observation, an upper limit on the count rate from \src\ was computed using the \textsc{ciao} tool \textsc{aprates} and taking into account the point-spread function fraction in the apertures within which the counts were extracted. In order to convert limits on the count rates from the different detectors into upper limits on the unabsorbed flux, we used the NASA/HEASARC \textsc{pimms} tool assuming a soft power-law spectrum with photon index $\Gamma\sim3$ and absorption $N_{\rm H}\sim7\times10^{21}$ cm$^{-2}$. The corresponding upper limits on the luminosity (assuming a distance of 780 kpc) range from $\sim$$6\times10^{35}$ to $5\times10^{37}$ \lum\ and are shown in Fig.~\ref{ulcxo}.

\begin{figure*}
\resizebox{\hsize}{!}{\includegraphics[angle=0]{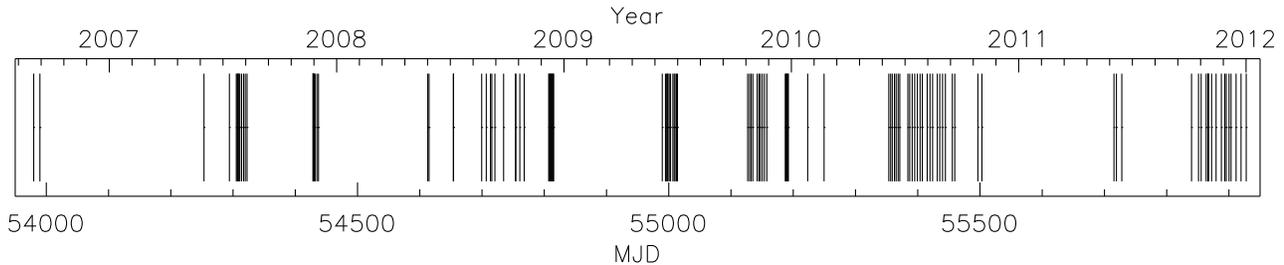}}
\caption{\label{archivalobss} Epochs of the \swift/XRT serendipitous observations of the position of \src.}
\end{figure*}

\subsection{\swift}\label{xrtul}
In the period between 2006 September 01 and 2012 January 01, \swift\ serendipitously imaged with the XRT (in PC mode) the position of \src\ 119 times for a total exposure of $\sim$370 ks. The spread of the  observations can be seen in Fig.~\ref{archivalobss} while the exposure time for each year is given in Table~\ref{ulimits}.  

We examined each observation, but the source was never detected, nor it was detected in the total and yearly total images. For each year and for the total data set, we computed 3$\sigma$ upper limits on the count rate (following \citealt*{kraft91}). These upper limits (Table~\ref{ulimits}) can be directly compared with the average (and fairly constant) PC rate observed in 2012 February--March, ($0.370\pm0.007$) counts s$^{-1}$. In particular, the last \swift\ observation performed before the discovery of \src\ (obs. ID: 00035336051; 2012 January 01) yields an upper limit of $2.5\times10^{-3}$ counts s$^{-1}$, implying for the source a flux increase of 150 or more during 2012 January/February.

\begin{table}
\centering
\caption{Pre-outburst upper limits from \swift/XRT serendipitous observations of the position of \src.}
\label{ulimits}
\begin{tabular}{@{}cccc}
\hline
Year & Observations & Total exposure & Count rate upper limit$^a$\\
 & & (ks) & (counts s$^{-1}$)\\
\hline
2006 & 2 & 9.9 & 0.0010 \\
2007 & 19 & 58.3 & 0.00038 \\
2008 & 23 & 65.3 & 0.00023 \\ 
2009 & 28 & 86.2 & 0.00028 \\
2010 & 27 & 98.5 & 0.00031 \\
2011 & 19 & 47.9 & 0.00066 \\
2012 & 1 & 4.1 & 0.0025 \\
\cline{1-3}
2006--2012 & 119 & 369.5 & 0.00016\\
\hline
\end{tabular}
\begin{list}{}{}
\item[$^{a}$] At a confidence level of 3$\sigma$.
\end{list}
\end{table}

\section{Discussion and Conclusions}\label{disc}

\src\ is a transient X-ray source in M31 that, at its maximum, reached luminosity in the ULX range. Undetected in all previous X-ray observations, in February 2012 it suddenly started to show powerful X-ray emission. After reaching the maximum, the source luminosity remained fairly constant at $\ga$$10^{39}$ \lum\ for at least $\sim$40 days, then it faded below $\approx$$10^{38}$ \lum\ in the following $\sim$200 days. The decay, accompanied by a spectral softening, can be described by a broken power-law with a break time of $\approx$100 days. No broad components nor narrow features (in the form of quasi-periodic oscillations) were detected in the power density spectra up to $\sim$280 Hz. We searched also for optical and near-UV emission from \src, but no no source was detected at its position down to a sensitivity limit of $\sim$22 in the $V$ and $B$ bands, and of 23--24 in the near UV.

In the following, we discuss the properties of \src\ only in the context of an accreting black hole in M31. The large absorption column and the lack of an optical counterpart exclude a foreground object. In terms of a background object, the best candidates are an active galactic nucleus or a tidal disruption event. However, no active galactic nucleus has been observed to vary in X-rays by more than two orders of magnitude, while a tidal disruption event is ruled out by the flux evolution, that does not follow the characteristic decay of such events \citep{burrows11short}. 

Our spectral analysis of the \swift\ PC data showed that the energy spectrum of \src\ can be fit using a disc component, and that the combination of a disk component and a power law  that is commonly used to describe both the spectra of some ULXs and of Galactic BH binaries is not statistically needed (see also \citealt*{barnard12}). However we cannot exclude the possibility that an additional component is not detected because of the low counting statistics. Along its outburst \src\ showed only moderate and slow spectral changes in contrast with what is usually seen in transient BH binaries in outburst (see e.g. \citealt*{motta09}). 

The transient ULXs observed so far appear to have rather different properties and \src\ is not an exception in this picture. The source presents similarities with other transients, but it cannot be definitely associated to any of them. 

M31 ULX-1 was observed to reach a peak luminosity of $5\times10^{39}$ \lum\ \citep{kaur12} followed by a decrement in flux with an $e$-folding time-scale of $\sim$40 days (\citealt{middleton12}). A blue optical counterpart to M31 ULX-1 was detected during the  outburst, but not in quiescence, pointing towards emission from an irradiated accretion disc. This suggests a low mass companion star transferring matter via Roche lobe overflow. The spectral properties of M31 ULX-1 are roughly described by a single disc component around a 10-$M_{\odot}$ BH with a spin of $a\sim0.4$.\footnote{The dimensionless spin parameter $a=cJ/(GM^2)$, where $J$ and $M$ are the angular momentum and mass of the black hole, respectively.} On the other hand, contrarily to what we observed in \src, the best description of the spectra is obtained with the addition of a second component which represents an optically thick medium (namely a low temperature corona  described by a \textsc{comptt}) close to the BH. It can be associated to the photosphere of a wind ejected by the inner regions of the disc, that is expected to set in at super-Eddington accretion rate. Indeed, the softer component of the spectrum can be also described by a slim disc model, indicating that advection may be important. The disc component becomes more important when the total luminosity decreases, suggesting that the wind unveils progressively the inner regions of disc when the accretion rate drops.

Also M83 ULX-1 \citep{soria12}  showed a long term evolution similar to \src\ but, as in the case of M31 ULX-1, the energy spectrum is significantly different from \src. Undetected ($L_{\mathrm{X}}<10^{36}$ \lum) before 2011, M83 ULX-1 was discovered by \textit{Chandra} and it was seen to reach a peak luminosity comparable to that of M31 \mbox{ULX-1} ($\sim$$4\times10^{39}$ \lum). Its spectral properties are modelled by a cold disc plus a power law without evidence of curvature at high energy, opposite to what is commonly seen in ULXs. In addition, the source does not show hints of a decline of the emission for at least 100 days after discovery and no spectral transitions were observed. Again contrarily to \src, a blue optical counterpart is observed, but only during the outburst, pointing to reprocessed emission from the outer accretion disc and the companion star, a red giant or AGB star with mass $M<4$ $M_{\odot}$ (\citealt{soria12}). 

Finally, the ULX in NGC5128, discovered by \textit{Chandra} in the 2009, was observed at a luminosity of $(2$--$3)\times10^{39}$ \lum. The outburst lasted for at least 70 days, but the source was no longer observed. It is an intriguing ULX because it experienced spectral state variations between a \textit{very high state}, dominated by a power law with a photon index higher than 2.2, and a \textit{high/soft} state, dominated by an accretion disc of $\sim$1 keV, consistent with the behaviour of the Galactic BH XRBs (\citealt{sivakoff08short}).

One possible interpretation of the spectra-timing properties of \src\ can be given by making an hypothesis on the nature of the source. If it is a stellar-mass BH transient, we can assume that most of the outburst evolution of the source could not be observed due to its distance from the observer. In this scenario, the source would be visible from Earth only during its brightest phases, which for many BH transients (e.g. H\,1743--322, XTE\,J1650--754, GRO\,J1655--40) is encountered during the soft spectral states \citep*{belloni11}. In the soft  states the energy spectrum is strongly dominated by a soft disk component (that sometimes is the only component visible in the X-ray spectrum) while the fast time variability is usually consistent with zero. These properties are consistent with what we reported on \src. The duration of this high-luminosity phase in a Galactic BH transient is variable depending on the source and on the properties of the single outburst. However, the average length of such periods (few months) is absolutely consistent with the duration of the outburst of \src\ (see e.g. the case of the Galactic BH transients XTE\,J1550--754, \citealt{kubota04} and GRO\,J1655--40, \citealt{motta12}). 

Assuming that at maximum luminosity \src\ was in a disc-dominated state and that it was radiating at a significant fraction of $L_{\mathrm{Edd}}$ (say $\sim$0.6), the mass of the BH would be $\sim$12 $M_\odot$. For small inclinations ($\leq$45$\degr$), a similar value is obtained from the normalisation of the disc component obtained from the fit of the combined pre-gap PC spectrum, assuming that the inner disc radius is truncated at 6 gravitational radii and that the disc spectrum has a standard color correction factor (see eq.~[2] in \citealt{zampieri09}; see also \citealt{lz09} and references therein). This is consistent with the hypothesis that \src\ could be indeed an accreting  stellar-mass black-hole binary observed in its soft state.

The upper limits in optical bands are sufficiently deep to place interesting constraints on the donor mass. Assuming no extinction and a distance modulus of 24.47 mag for M31 (from the NASA/IPAC Extragalactic Database),\footnote{See http://ned.ipac.caltech.edu/.} the upper limit in the $V$ band translates into an upper limit on the absolute magnitude $M_V>-2.8$. Taking binary evolution effects and X-ray irradiation into account, a stellar-mass BH accreting through Roche lobe overflow is consistent with this upper limit if the donor is a main sequence star of 8--10 $M_\odot$ or a giant of $< $8 $M_\odot$ \citep{pz10}. In fact, for a donor below 5 $M_\odot$, the disc is no longer stable \citep{dubus99,pz08}, in agreement with the transient nature of the source.

\section*{Acknowledgments} 
PE wishes to dedicate this paper to his mother for all intents and purposes, Maria Pastorcich; she died on March 24, 2012, but her spirit and smile live on in everyone who knew her. This research is based on observations with the NASA/UKSA/ASI mission \swift\ and with the Copernico Telescope at Cima Ekar, Asiago (Italy), operated by INAF--Astronomical Observatory of Padua. We also used data and software provided by the \cxo\ X-ray Center (CXC, operated for NASA by SAO) and the ESA's \xmm\ Science Archive (XSA). We thank Anna Wolter for useful comments on the manuscript, Fabio Pizzolato for helpful discussions, and the referee, Knox Long, for valuable comments. This work was partially supported by the Italian Space Agency through ASI--INAF contracts I/009/10/0 and I/004/11/0. FP and LZ acknowledge financial support from INAF through grant PRIN-2011-1.

\bibliographystyle{mn2e}
\bibliography{biblio}

\begin{thebibliography}{}

\bibitem[\protect\citeauthoryear{{Anders} \& {Grevesse}}{{Anders} \&
  {Grevesse}}{1989}]{anders89}
{Anders} E.,  {Grevesse} N.,  1989, \gca, 53, 197

\bibitem[\protect\citeauthoryear{{Arnaud}}{{Arnaud}}{1996}]{arnaud96}
{Arnaud} K.~A.,  1996, in {Jacoby}, G.~H. and {Barnes}, J., eds., Astronomical
  Data Analysis Software and Systems V. Vol.~101 of ASP Conf. Ser., San
  Francisco CA,
p.~17

\bibitem[\protect\citeauthoryear{{Balucinska-Church} \&
  {McCammon}}{{Balucinska-Church} \& {McCammon}}{1992}]{balucinska92}
{Balucinska-Church} M.,  {McCammon} D.,  1992, \apj, 400, 699

\bibitem[\protect\citeauthoryear{{Barnard}, {Garcia} \& {Murray}}{{Barnard}
  et~al.}{2012}]{barnard12}
{Barnard} R.,  {Garcia} M.~R.,    {Murray} S.~S.,  2012, Astron. Tel., 3937

\bibitem[\protect\citeauthoryear{{Barthelmy} et~al.,}{{Barthelmy}
  et~al.}{2005}]{barthelmy05short}
{Barthelmy} S.~D.  et~al., 2005, Space Science Reviews, 120, 143

\bibitem[\protect\citeauthoryear{{Begelman}}{{Begelman}}{2002}]{begelman02}
{Begelman} M.~C.,  2002, \apjl, 568, L97

\bibitem[\protect\citeauthoryear{{Begelman}, {King} \& {Pringle}}{{Begelman}
  et~al.}{2006}]{begelman06}
{Begelman} M.~C.,  {King} A.~R.,    {Pringle} J.~E.,  2006, \mnras, 370, 399

\bibitem[\protect\citeauthoryear{{Belczynski}, {Bulik}, {Fryer}, {Ruiter},
  {Valsecchi}, {Vink} \& {Hurley}}{{Belczynski} et~al.}{2010}]{belczynski10}
{Belczynski} K.,  {Bulik} T.,  {Fryer} C.~L.,  {Ruiter} A.,  {Valsecchi} F.,
  {Vink} J.~S.,    {Hurley} J.~R.,  2010, \apj, 714, 1217

\bibitem[\protect\citeauthoryear{{Belloni}, {Motta} \&
  {Mu{\~n}oz-Darias}}{{Belloni} et~al.}{2011}]{belloni11}
{Belloni} T.~M.,  {Motta} S.~E.,    {Mu{\~n}oz-Darias} T.,  2011, Bulletin of
  the Astronomical Society of India, 39, 409

\bibitem[\protect\citeauthoryear{{Breeveld}, {Landsman}, {Holland}, {Roming},
  {Kuin} \& {Page}}{{Breeveld} et~al.}{2011}]{breeveld11}
{Breeveld} A.~A.,  {Landsman} W.,  {Holland} S.~T.,  {Roming} P.,  {Kuin}
  N.~P.~M.,    {Page} M.~J.,  2011, in {J.~E.~McEnery, J.~L.~Racusin, 
  N.~Gehrels} ed., Gamma Ray Bursts 2010. Vol.~1358 of AIP Conf. Proc.,
  Melville NY, p. 373

\bibitem[\protect\citeauthoryear{{Burrows} et~al.,}{{Burrows}
  et~al.}{2005}]{burrows05short}
{Burrows} D.~N.  et~al., 2005, Space Science Reviews, 120, 165

\bibitem[\protect\citeauthoryear{{Burrows} et~al.,}{{Burrows}
  et~al.}{2011}]{burrows11short}
{Burrows} D.~N.  et~al., 2011, \nat, 476, 421

\bibitem[\protect\citeauthoryear{{Casella}, {Belloni} \& {Stella}}{{Casella}
  et~al.}{2005}]{casella05}
{Casella} P.,  {Belloni} T.,    {Stella} L.,  2005, \apj, 629, 403

\bibitem[\protect\citeauthoryear{{Colbert} \& {Mushotzky}}{{Colbert} \&
  {Mushotzky}}{1999}]{colbert99}
{Colbert} E.~J.~M.,  {Mushotzky} R.~F.,  1999, \apj, 519, 89

\bibitem[\protect\citeauthoryear{{Davis}, {Narayan}, {Zhu}, {Barret},
  {Farrell}, {Godet}, {Servillat} \& {Webb}}{{Davis} et~al.}{2011}]{davis11}
{Davis} S.~W.,  {Narayan} R.,  {Zhu} Y.,  {Barret} D.,  {Farrell} S.~A.,
  {Godet} O.,  {Servillat} M.,    {Webb} N.~A.,  2011, \apj, 734, 111

\bibitem[\protect\citeauthoryear{{Dubus}, {Lasota}, {Hameury} \&
  {Charles}}{{Dubus} et~al.}{1999}]{dubus99}
{Dubus} G.,  {Lasota} J.-P.,  {Hameury} J.-M.,    {Charles} P.,  1999, \mnras,
  303, 139

\bibitem[\protect\citeauthoryear{{Fabbiano}}{{Fabbiano}}{1989}]{fabbiano89}
{Fabbiano} G.,  1989, \araa, 27, 87

\bibitem[\protect\citeauthoryear{{Farrell}, {Webb}, {Barret}, {Godet} \&
  {Rodrigues}}{{Farrell} et~al.}{2009}]{farrell09}
{Farrell} S.~A.,  {Webb} N.~A.,  {Barret} D.,  {Godet} O.,    {Rodrigues}
  J.~M.,  2009, \nat, 460, 73

\bibitem[\protect\citeauthoryear{{Feng}, {Rao} \& {Kaaret}}{{Feng}
  et~al.}{2010}]{feng10}
{Feng} H.,  {Rao} F.,    {Kaaret} P.,  2010, \apjl, 710, L137

\bibitem[\protect\citeauthoryear{{Feng} \& {Soria}}{{Feng} \&
  {Soria}}{2011}]{feng11}
{Feng} H.,  {Soria} R.,  2011, \nar, 55, 166

\bibitem[\protect\citeauthoryear{{Fitzpatrick} \& {Massa}}{{Fitzpatrick} \&
  {Massa}}{2007}]{fitzpatrick07}
{Fitzpatrick} E.~L.,  {Massa} D.,  2007, \apj, 663, 320

\bibitem[\protect\citeauthoryear{{Garmire}, {Bautz}, {Ford}, {Nousek} \&
  {Ricker} Jr.}{{Garmire} et~al.}{2003}]{garmire03}
{Garmire} G.~P.,  {Bautz} M.~W.,  {Ford} P.~G.,  {Nousek} J.~A.,    {Ricker}
  Jr. G.~R.,  2003, in {Truemper}, J.~E. and {Tananbaum}, H.~D., eds., X-Ray 
and Gamma-Ray Telescopes and Instruments for
  Astronomy, Vol.~4851 of
  Proc. SPIE. SPIE, Bellingham WA,
p. 28

\bibitem[\protect\citeauthoryear{{Gladstone}, {Roberts} \& {Done}}{{Gladstone}
  et~al.}{2009}]{gladstone09}
{Gladstone} J.~C.,  {Roberts} T.~P.,    {Done} C.,  2009, \mnras, 397, 1836

\bibitem[\protect\citeauthoryear{{Helfand}}{{Helfand}}{1984}]{h84}
{Helfand} D.~J.,  1984, \pasp, 96, 913

\bibitem[\protect\citeauthoryear{{Henze}, {Pietsch} \& {Haberl}}{{Henze}
  et~al.}{2012a}]{hepiha12}
{Henze} M.,  {Pietsch} W.,    {Haberl} F.,  2012a, Astron. Tel., 3959

\bibitem[\protect\citeauthoryear{{Henze}, {Pietsch} \& {Haberl}}{{Henze}
  et~al.}{2012b}]{henze12}
{Henze} M.,  {Pietsch} W.,    {Haberl} F.,  2012b, Astron. Tel., 3890

\bibitem[\protect\citeauthoryear{{Henze}, {Pietsch} \& {Haberl}}{{Henze}
  et~al.}{2012c}]{hph12}
{Henze} M.,  {Pietsch} W.,    {Haberl} F.,  2012c, Astron. Tel., 3921

\bibitem[\protect\citeauthoryear{{Henze}, {Pietsch}, {Haberl} \&
  {Greiner}}{{Henze} et~al.}{2012}]{hphg12}
{Henze} M.,  {Pietsch} W.,  {Haberl} F.,    {Greiner} J.,  2012, Astron. Tel.,
  4125

\bibitem[\protect\citeauthoryear{{Holland}}{{Holland}}{1998}]{holland98}
{Holland} S.,  1998, \aj, 115, 1916

\bibitem[\protect\citeauthoryear{{Kaur}, {Henze}, {Haberl}, {Pietsch},
  {Greiner}, {Rau}, {Hartmann}, {Sala} \& {Hernanz}}{{Kaur}
  et~al.}{2012}]{kaur12}
{Kaur} A.,  {Henze} M.,  {Haberl} F.,  {Pietsch} W.,  {Greiner} J.,  {Rau} A.,
  {Hartmann} D.~H.,  {Sala} G.,    {Hernanz} M.,  2012, \aap, 538, A49

\bibitem[\protect\citeauthoryear{{King}, {Davies}, {Ward}, {Fabbiano} \&
  {Elvis}}{{King} et~al.}{2001}]{king01}
{King} A.~R.,  {Davies} M.~B.,  {Ward} M.~J.,  {Fabbiano} G.,    {Elvis} M.,
  2001, \apjl, 552, L109

\bibitem[\protect\citeauthoryear{{Kraft}, {Burrows} \& {Nousek}}{{Kraft}
  et~al.}{1991}]{kraft91}
{Kraft} R.~P.,  {Burrows} D.~N.,    {Nousek} J.~A.,  1991, \apj, 374, 344

\bibitem[\protect\citeauthoryear{{Kubota} \& {Done}}{{Kubota} \&
  {Done}}{2004}]{kubota04}
{Kubota} A.,  {Done} C.,  2004, \mnras, 353, 980

\bibitem[\protect\citeauthoryear{{Leahy}, {Darbro}, {Elsner}, {Weisskopf},
  {Kahn}, {Sutherland} \& {Grindlay}}{{Leahy} et~al.}{1983}]{leahy83}
{Leahy} D.~A.,  {Darbro} W.,  {Elsner} R.~F.,  {Weisskopf} M.~C.,  {Kahn} S.,
  {Sutherland} P.~G.,    {Grindlay} J.~E.,  1983, \apj, 266, 160

\bibitem[\protect\citeauthoryear{{Long} \& {van Speybroeck}}{{Long} \& {van
  Speybroeck}}{1983}]{long83}
{Long} K.~S.,  {van Speybroeck} L.~P.,  1983, in {Lewin} W.~H.~G.,  {van den
  Heuvel} E.~P.~J.,  eds., Accretion-Driven Stellar X-ray Sources, Cambridge
  University Press, Cambridge,
p.~141

\bibitem[\protect\citeauthoryear{{Lorenzin} \& {Zampieri}}{{Lorenzin} \& {Zampieri}}{2009}]{lz09} {Lorenzin} A., {Zampieri} L., 2009, \mnras, 394, 1588

\bibitem[\protect\citeauthoryear{{Magnier}, {Lewin}, {van Paradijs},
  {Hasinger}, {Jain}, {Pietsch} \& {Truemper}}{{Magnier}
  et~al.}{1992}]{magnier92}
{Magnier} E.~A.,  {Lewin} W.~H.~G.,  {van Paradijs} J.,  {Hasinger} G.,  {Jain}
  A.,  {Pietsch} W.,    {Truemper} J.,  1992, \aaps, 96, 379

\bibitem[\protect\citeauthoryear{{Mapelli}, {Colpi} \& {Zampieri}}{{Mapelli}
  et~al.}{2009}]{mapelli09}
{Mapelli} M.,  {Colpi} M.,    {Zampieri} L.,  2009, \mnras, 395, L71

\bibitem[\protect\citeauthoryear{{Middleton}, {Sutton} \&
  {Roberts}}{{Middleton} et~al.}{2011}]{middleton11}
{Middleton} M.~J.,  {Sutton} A.~D.,    {Roberts} T.~P.,  2011, \mnras, 417, 464

\bibitem[\protect\citeauthoryear{{Middleton}, {Sutton}, {Roberts}, {Jackson} \&
  {Done}}{{Middleton} et~al.}{2012}]{middleton12}
{Middleton} M.~J.,  {Sutton} A.~D.,  {Roberts} T.~P.,  {Jackson} F.~E.,
  {Done} C.,  2012, \mnras, 420, 2969

\bibitem[\protect\citeauthoryear{{Motta}, {Belloni} \& {Homan}}{{Motta}
  et~al.}{2009}]{motta09}
{Motta} S.,  {Belloni} T.,    {Homan} J.,  2009, \mnras, 400, 1603

\bibitem[\protect\citeauthoryear{{Motta}, {Homan}, {Mu\~{n}oz-Darias},
  {Casella}, {Belloni}, {Hiemstra} \& {M\`endez, M}.}{{Motta}
  et~al.}{2012}]{motta12}
{Motta} S.,  {Homan} J.,  {Mu\~{n}oz-Darias} T.,  {Casella} P.,  {Belloni}
  T.~M.,  {Hiemstra} B.,    {M\`endez, M}. 2012, \mnras, in press (eprint:
  astro-ph.HE/1209.0327)

\bibitem[\protect\citeauthoryear{{Mucciarelli}, {Casella}, {Belloni},
  {Zampieri} \& {Ranalli}}{{Mucciarelli} et~al.}{2006}]{mucciarelli06}
{Mucciarelli} P.,  {Casella} P.,  {Belloni} T.,  {Zampieri} L.,    {Ranalli}
  P.,  2006, \mnras, 365, 1123

\bibitem[\protect\citeauthoryear{{Murray}, {Austin}, {Chappell}, {Gomes},
  {Kenter}, {Kraft}, {Meehan}, {Zombeck}, {Fraser} \& {Serio}}{{Murray}
  et~al.}{2000}]{murray00}
{Murray} S.~S. et~al.,  2000, in {Truemper} J.~E.,  {Aschenbach} B.,  eds., X-Ray 
Optics, Instruments, and Missions III, Vol. 4012 of Proc. SPIE. SPIE, Bellingham WA, 
p.~68

\bibitem[\protect\citeauthoryear{{Oke} \& {Gunn}}{{Oke} \&
  {Gunn}}{1983}]{oke83}
{Oke} J.~B.,  {Gunn} J.~E.,  1983, \apj, 266, 713

\bibitem[\protect\citeauthoryear{{Patruno} \& {Zampieri}}{{Patruno} \&
  {Zampieri}}{2008}]{pz08}
{Patruno} A.,  {Zampieri} L.,  2008, \mnras, 386, 543

\bibitem[\protect\citeauthoryear{{Patruno} \& {Zampieri}}{{Patruno} \&
  {Zampieri}}{2010}]{pz10}
{Patruno} A.,  {Zampieri} L.,  2010, \mnras, 403, L69

\bibitem[\protect\citeauthoryear{{Pintore} \& {Zampieri}}{{Pintore} \&
  {Zampieri}}{2012}]{pintore12}
{Pintore} F.,  {Zampieri} L.,  2012, \mnras, 420, 1107

\bibitem[\protect\citeauthoryear{{Poole} et~al.,}{{Poole}
  et~al.}{2008}]{poole08short}
{Poole} T.~S.  et~al., 2008, \mnras, 383, 627

\bibitem[\protect\citeauthoryear{{Predehl} \& {Schmitt}}{{Predehl} \&
  {Schmitt}}{1995}]{predehl95}
{Predehl} P.,  {Schmitt} J.~H.~M.~M.,  1995, \aap, 293, 889

\bibitem[\protect\citeauthoryear{{Roberts}}{{Roberts}}{2007}]{roberts07}
{Roberts} T.~P.,  2007, \apss, 311, 203

\bibitem[\protect\citeauthoryear{{Roming} et~al.,}{{Roming}
  et~al.}{2005}]{roming05short}
{Roming} P.~W.~A.  et~al., 2005, Space Science Reviews, 120, 95

\bibitem[\protect\citeauthoryear{{Schlegel}, {Finkbeiner} \&
  {Davis}}{{Schlegel} et~al.}{1998}]{schlegel98}
{Schlegel} D.~J.,  {Finkbeiner} D.~P.,    {Davis} M.,  1998, \apj, 500, 525

\bibitem[\protect\citeauthoryear{{Shaposhnikov} \& {Titarchuk}}{{Shaposhnikov}
  \& {Titarchuk}}{2009}]{shaposhnikov09}
{Shaposhnikov} N.,  {Titarchuk} L.,  2009, \apj, 699, 453

\bibitem[\protect\citeauthoryear{{Sivakoff} et~al.,}{{Sivakoff}
  et~al.}{2008}]{sivakoff08short}
{Sivakoff} G.~R.  et~al., 2008, \apjl, 677, L27

\bibitem[\protect\citeauthoryear{{Soria}, {Kuntz}, {Winkler}, {Blair}, {Long},
  {Plucinsky} \& {Whitmore}}{{Soria} et~al.}{2012}]{soria12}
{Soria} R.,  {Kuntz} K.~D.,  {Winkler} P.~F.,  {Blair} W.~P.,  {Long} K.~S.,
  {Plucinsky} P.~P.,    {Whitmore} B.~C.,  2012, \apj, 750, 152

\bibitem[\protect\citeauthoryear{{Stanek} \& {Garnavich}}{{Stanek} \&
  {Garnavich}}{1998}]{stanek98}
{Stanek} K.~Z.,  {Garnavich} P.~M.,  1998, \apjl, 503, L131

\bibitem[\protect\citeauthoryear{{Stobbart}, {Roberts} \& {Wilms}}{{Stobbart}
  et~al.}{2006}]{stobbart06}
{Stobbart} A.-M.,  {Roberts} T.~P.,    {Wilms} J.,  2006, \mnras, 368, 397

\bibitem[\protect\citeauthoryear{{Strohmayer} \& {Mushotzky}}{{Strohmayer} \&
  {Mushotzky}}{2003}]{sm03}
{Strohmayer} T.~E.,  {Mushotzky} R.~F.,  2003, \apjl, 586, L61

\bibitem[\protect\citeauthoryear{{Strohmayer} \& {Mushotzky}}{{Strohmayer} \&
  {Mushotzky}}{2009}]{strohmayer09}
{Strohmayer} T.~E.,  {Mushotzky} R.~F.,  2009, \apj, 703, 1386

\bibitem[\protect\citeauthoryear{{Strohmayer}, {Mushotzky}, {Winter}, {Soria},
  {Uttley} \& {Cropper}}{{Strohmayer} et~al.}{2007}]{strohmayer07}
{Strohmayer} T.~E.,  {Mushotzky} R.~F.,  {Winter} L.,  {Soria} R.,  {Uttley}
  P.,    {Cropper} M.,  2007, \apj, 660, 580

\bibitem[\protect\citeauthoryear{{Zacharias}, {Monet}, {Levine}, {Urban},
  {Gaume} \& {Wycoff}}{{Zacharias} et~al.}{2005}]{zacharias05}
{Zacharias} N.,  {Monet} D.~G.,  {Levine} S.~E.,  {Urban} S.~E.,  {Gaume} R.,
   {Wycoff} G.~L.,  2005, VizieR Online Data Catalog, 1297, 0

\bibitem[\protect\citeauthoryear{{Zampieri} \& {Roberts}}{{Zampieri} \&
  {Roberts}}{2009}]{zampieri09}
{Zampieri} L.,  {Roberts} T.~P.,  2009, \mnras, 400, 677

\end{thebibliography}

\bsp

\label{lastpage}

\end{document}